\documentclass[conference]{IEEEtran}
\usepackage{cite}
\usepackage{amsmath,amssymb,amsfonts}
\usepackage{algorithmic}
\usepackage{graphicx}
\usepackage{textcomp}
\usepackage{xcolor}
\usepackage{cite}
\usepackage{algorithmic}
\usepackage{algorithm}
\usepackage{enumerate}
\usepackage{subcaption}
\def\BibTeX{{\rm B\kern-.05em{\sc i\kern-.025em b}\kern-.08em
    T\kern-.1667em\lower.7ex\hbox{E}\kern-.125emX}}
\begin{document}

\title{Beamforming in Wireless Coded-Caching
Systems\\
{\footnotesize \textsuperscript{}}
\thanks{}
}

\author{\IEEEauthorblockN{1\textsuperscript{st} Sneha Madhusudan}
\IEEEauthorblockA{
\textit{Ericsson AB}\\
Stockholm, Sweden \\
sneha.madhusudan@ericsson.com}

\and
\IEEEauthorblockN{2\textsuperscript{nd} Charitha Madapatha}
\IEEEauthorblockA{\textit{Dept. of Electrical Engineering
} \\
\textit{Chalmers University of Technology}\\
Gothenburg, Sweden \\
charitha@chalmers.se}
\and
\IEEEauthorblockN{3\textsuperscript{rd} Behrooz Makki}
\IEEEauthorblockA{\textit{Ericsson Research} \\
\textit{Ericsson AB}\\
Gothenburg, Sweden \\
behrooz.makki@ericsson.com}
\and

\IEEEauthorblockN{4\textsuperscript{th} Hao Guo}
\IEEEauthorblockA{\textit{Dept. of Electrical Engineering} \\
\textit{Chalmers University of Technology}\\
Gothenburg, Sweden \\
hao.guo@chalmers.se}
\and
\IEEEauthorblockN{5\textsuperscript{th} Tommy Svensson}
\IEEEauthorblockA{\textit{Dept. of Electrical Engineering} \\
\textit{Chalmers University of Technology}\\
Gothenburg, Sweden \\
tommy.svensson@chalmers.se}

}

\maketitle

\begin{abstract}

\textcolor{black}{Increased capacity in the access network poses capacity challenges on the transport network due to the aggregated traffic. However, there are spatial and time correlation in the user data demands that could potentially be utilized. To that end, we investigate a wireless transport network architecture that integrates beamforming and coded-caching strategies.} Especially, our proposed design entails a server with multiple antennas that broadcasts content to cache nodes responsible for serving users. \textcolor{black}{Traditional caching methods face the limitation of relying on the individual memory with additional overhead. Hence, we develop an efficient
genetic algorithm-based scheme for beam optimization in the coded-caching system.} By exploiting the advantages of beamforming and coded-caching, the architecture achieves gains in terms of multicast opportunities, interference mitigation, and reduced peak backhaul traffic. A comparative analysis of this joint design with traditional, un-coded caching schemes is also conducted to assess the benefits of the proposed approach. Additionally, we examine the impact of various buffering and decoding methods on the performance of the coded-caching scheme. Our findings suggest that proper beamforming is useful in enhancing the effectiveness of the coded-caching technique, resulting in significant reduction in peak backhaul traffic.
\end{abstract}


\section{Introduction}
The ever-increasing data traffic and users' demands for high rates and reliable connections in 5G and beyond (6G) have necessitated network densification through the deployment of multiple base stations (BSs) of various types. However, the additional BSs require connectivity to the core network through the transport network \cite{mesodiakaki20236g}. The surge in traffic between the BS and core network, referred to as backhaul traffic, can result in backhaul congestion and ultimately lead to excessive network end-to-end (E2E) delays. To mitigate this issue and minimize backhaul load and E2E latency, wireless caching schemes have been proposed as a solution \cite{7565183,9475988}.

The process of storing most frequently accessed end-user contents closer to end-users in order to ease the backhaul load is referred to as caching. Here, during the low traffic demanding periods, the contents are placed near to the end-users. Caching has variety of use cases, and is especially useful in \textcolor{black}{delay-contrained communication where the data needs among the users are correlated in a confined space and time. Such use cases can be found in integrated access and backhaul (IAB) \cite{madapatha2020integrated,10119093,8891507,9677384}, vehicle-to-everything (V2X) \cite{8830373}, and  device-to-device (D2D) communication scenarios \cite{7150324}}. The technique has promising applicability in alleviating the effect on backhaul link rates during periods with high traffic demand. 
Coded caching (CC) introduced in \cite{Maddah-Ali, 6807823, 7537173}, is an innovative technique used in wireless comminication to optimize the data delivery and reduce the congestion. CC focuses on using all memories in the network when users request files during the delivery phase (DP), and access the cache node contents already placed during the placement phase (PP), resulting in lower peak \textcolor{black}{backhaul} traffic. CC has been previoulsy evaluated for various aspects, including the effect of delayed channel state information at the transmitter (CSIT), and wireless channel with mixed CSIT \cite{7857805,7864374,7962818}. Interference management \cite{8624603}, and rate splitting \cite{8007039} in wireless CC networks have also been studied. 

Depending on the system requirements, several strategies can be used at the \textcolor{black}{receiver} end to decode the received signals (See Section \ref{systemmodel} for more details).
In traditional caching, the efficiency gains are lower \textcolor{black}{compared to CC}, when the cache memory at each node is insufficient to store the needed amounts of popular content. CC, however, overcomes the limitations of traditional caching as the system is not solely relying on the individual memory at each node.

With the increased number of user equipments (UEs) using various applications, the mobile networks demand techniques that increase the data rates with lower level of interference. Beamforming is a tehnique which concentrates the signal in specific directions in order to maximize the signal quality, energy efficiency and achievable data rates \cite{9573748}. \textcolor{black}{Variety of beamforming methods are proposed for the evolving 5G and upcoming 6G systems in access and backhaul \cite{8371237,7959905}. A genetic algorithm (GA), \cite{madapatha2021topology}, based beamformer optimization approach for millimeter wave communication is investigated in \cite{7959905}, \cite{makki2016genetic}. Content delivery problems in
fading multi-input single-output (MISO) channels is studied in \cite{8094982}. Furthermore, CC with zero-forcing (ZF) is studied in \cite{8006902}. In \cite{8437354} it is shown that optimization of beamformers with UE-side CC in the access can lead to substaintial performance gains. However, to our knowledge there is no work reported on beamformed CC in wireless transport networks.}

\textcolor{black}{In this paper, we use beamforming along with CC techniques in order to improve the efficiency and data delivery speed in the wireless transport network by developing a novel GA-based scheme for beam optimization in the CC system. Thereby, we study the effect of beamforming on its efficiency, compared to uncoded caching schemes.} We present and evaluate the effect of different parameters and message decoding/buffering methods on the network performance, in terms of successful transmission probability (STP) and throughput. As we show, by using CC and beamforming, the backhaul traffic can be minimized and the STP can be maximized. \textcolor{black}{Moreover, we show that different buffering/decoding schemes at the cache nodes lead to different STP/throughput with different levels of complexity.}

\section{System Model}
\label{systemmodel}

\begin{figure}
\centerline{\includegraphics[width=2.5in]{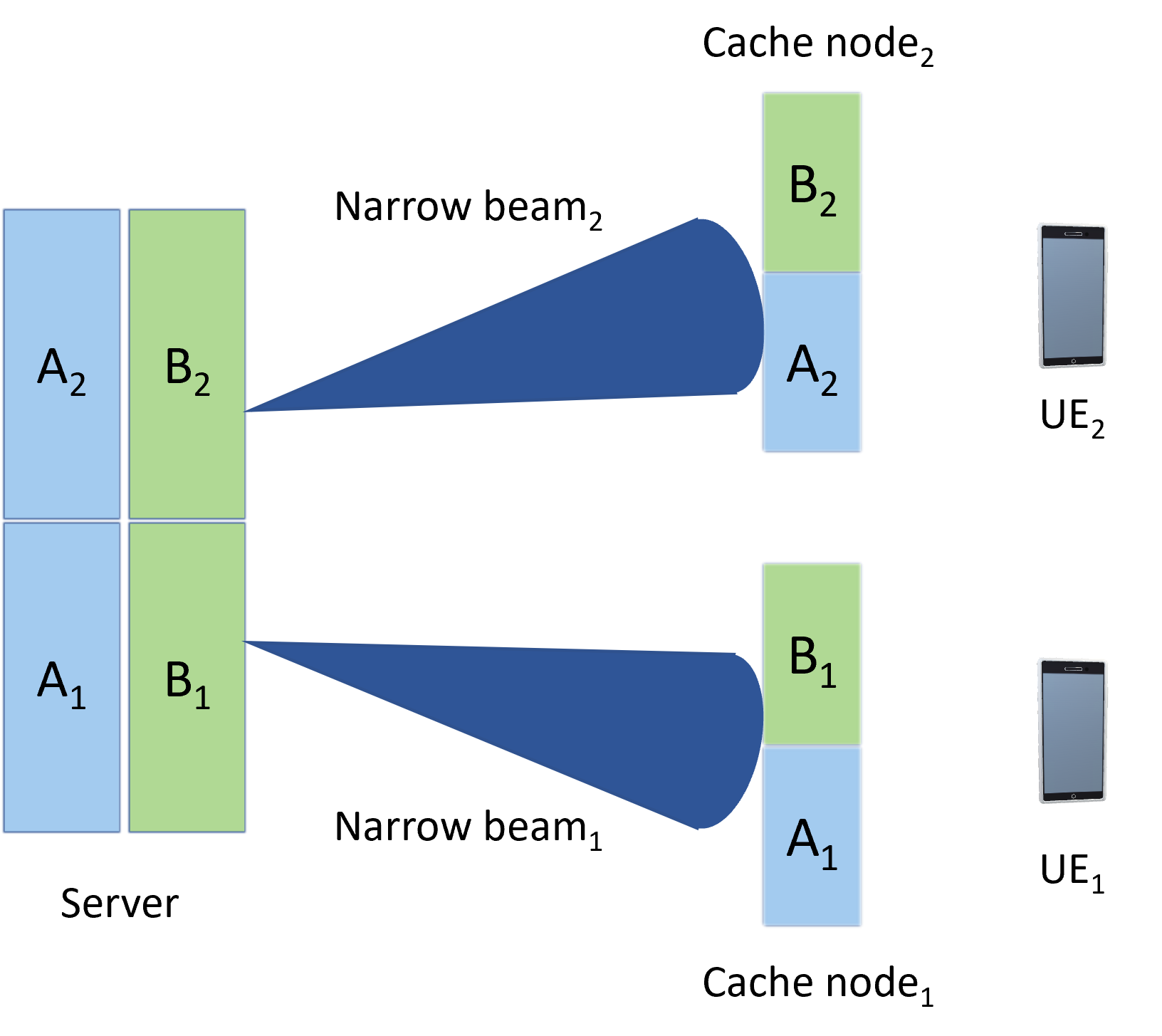}}
\caption{Placement phase of beamformed coded caching.}
\label{placementphase}
\end{figure}

\begin{figure}
\centerline{\includegraphics[width=3.5in]{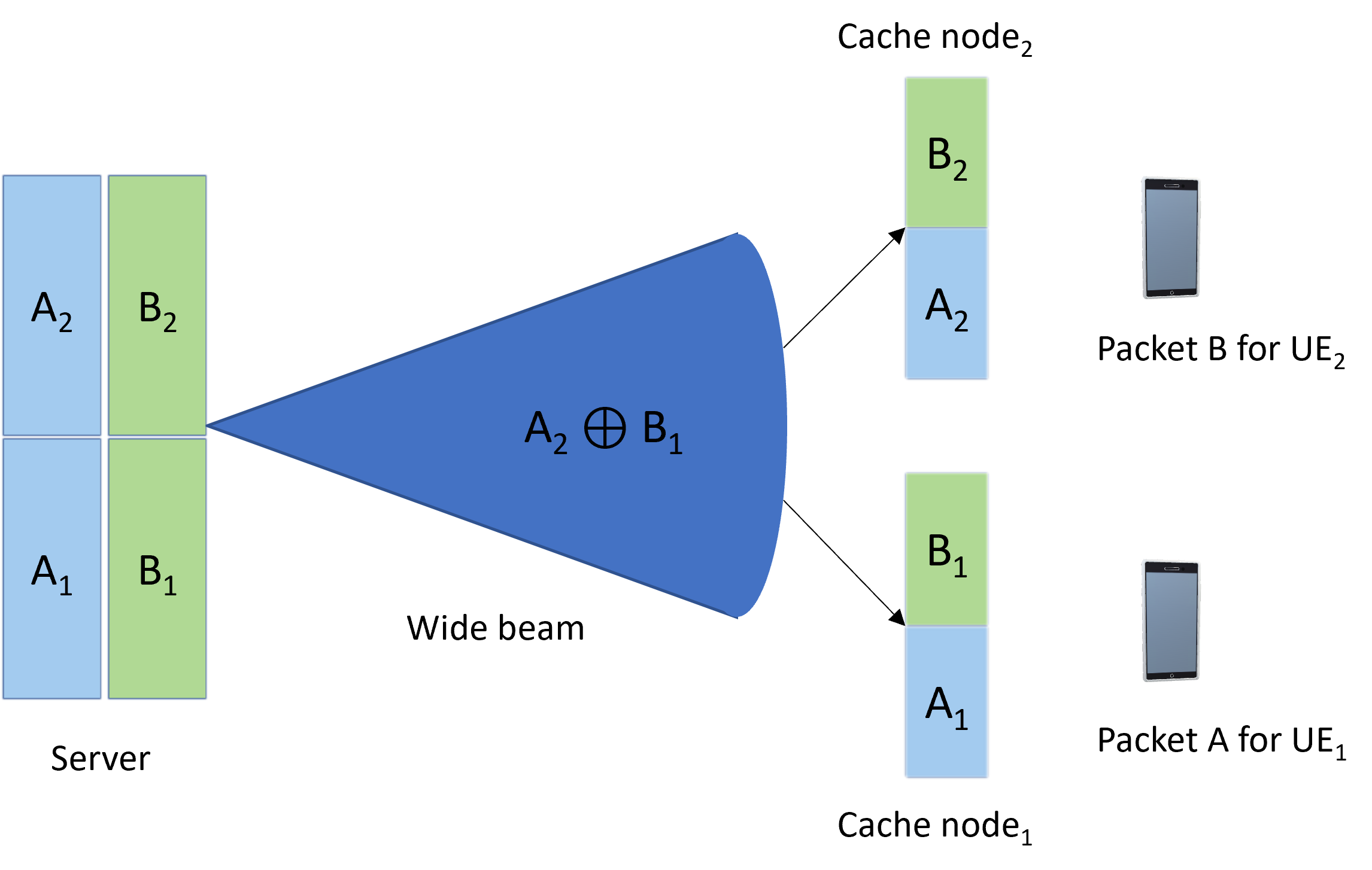}}
\caption{Delivery phase of beamformed coded caching.}
\label{deliveryphase}
\end{figure}

Beamforming is used along with the CC system in order to route the packets to the intended UEs which eventually increases the signal quality ensuring efficient decoding. Here, in the CC scheme, the broadcasted encoded signals are accessible and can be observed by all designated users\cite{8437354}. 

As an illustrative example, consider a mechanism in a network having an $L$-antenna server with $N = 2$ equally popular files, A, B, and $K = 2$ single-antenna cache nodes. Consider that each \textcolor{black}{cache node} can cache up to $M = 1$ file during the low traffic PP and each user requests for a single file during the high traffic DP. The files in the server constitute of two parts, designated as $A_1$, $A_2$ and $B_1$, $B_2$, respectively. This process can be divided into two main stages, namely, PP and DP.

\begin{itemize}
    \item PP: In this phase, the cache nodes receive different portions of files from the server using distinct narrow beams. The beams are used in different time slots to direct the sub-packets to appropriate cache nodes. Figure \ref{placementphase} illustrates the mechanism of data \textcolor{black}{placement} for the cache nodes. This intended data distribution is achieved through the beamforming vectors generating narrow beams. 
    In particular, the signals received at each of the cache nodes during this PP are given by



\begin{equation} 
R_{C_1}(t) = \sqrt{P} \boldsymbol{h_{1}^{\text{PP}} v_{1}} \begin{bmatrix} A_{1}(t) & B_{1}(t) \end{bmatrix} + \epsilon_{C_{1,1}}(t) \epsilon_{C_{1,2}}(t),
\label{eq1}
\end{equation}

\begin{equation}
R_{C_2}(t) = \sqrt{P} \boldsymbol{h_{2}^{\text{PP}} v_{2}} \begin{bmatrix} A_{2}(t) & B_{2}(t) \end{bmatrix} + \epsilon_{C_{2,1}}(t).
 \epsilon_{C_{2,2}}(t),
\end{equation}

\textcolor{black}{Herein, $P$ represents the total transmit power of the server. The vectors $\boldsymbol{h_{i}^{\text{PP}}}\in{\mathbb{C}^{N_{T} \times 1}}$, where $i = 1, 2$, denote the channel realizations during the PP for different caching nodes, and  $N_T$ denotes the number of transmit antennas. The vectors $\boldsymbol{v_{i}}\in{\mathbb{C}^{N_{T} \times 1}}$ where $i = 1, 2$, signify the beamforming vectors used to generate narrow beams for placing sub-files at the corresponding caching nodes. Moreover, $\epsilon_{C_{i,j}}$, $i, j = 1, 2$, represents the additive Gaussian noise, possessing a unit variance. Note that by the $\begin{bmatrix} A_{1}(t) & B_{1}(t) \end{bmatrix}$ representation, we denote the concatenation of $A_1 \in \mathbb{C}^{M \times 1}$ and $B_1 \in \mathbb{C}^{M \times 1}$.} Thereby, if $A_1, A_2, B_1, B_2$ are of length $M$, then, $R_{C_1}$, $R_{C_2}$ are of length $2M$.

\item DP: During the DP characterized by high traffic, the data requested are broadcasted to the users optimally to minimize the total delivery time. A wide, shared beam is used to broadcast the universally coded signal to the cache nodes. Figure \ref{deliveryphase} portrays the data broadcasting mechanism during this phase. Assuming the network comprises $F = 2$ files and $K = 2$ users, without loss of generality, the transmitted signals during the DP can be described as
\textcolor{black}{
\begin{equation}
R_{C_1}(t) = \sqrt{P} \boldsymbol{h_{1}^{\text{DP}} v_{1,2}} Y(t) + \epsilon_{C_{1,1}}(t),
\end{equation}
\begin{equation}
R_{C_2}(t) = \sqrt{P} \boldsymbol{h_{2}^{\text{DP}} v_{1,2}} Y(t) + \epsilon_{C_{2,2}}(t).
\label{eq4}
\end{equation}}
\textcolor{black}{In these expressions, $\boldsymbol{h_{i}^{\text{DP}}}\in{\mathbb{C}^{N_{T} \times 1}}$ with $i = 1, 2$, refers to the channel realization instances during the DP for individual users. Furthermore, the common coded signal $Y(t)$ is given by}
\begin{equation}
    Y(t) = \beta A_{2}(t) + \sqrt{1- \beta^{2}} B_{1}(t),
\label{eqs}    
\end{equation}
where, $\beta \in [0,1]$ is the power split parameter of the sub-files that will be used for optimization. \textcolor{black}{That is, $Y(y)$ is the superimposed signal of the sub-files $A_2$ and $B_1$, which are superimposed with different weights such that the total transmit power of the server is still $P$}.
The vector $\boldsymbol{v_{1,2}}$ refers to the beamforming vector used to create a shared wide beam that aids in transmitting the uniformly coded signal. The effectiveness and performance of the DP are directly influenced by the preceding PP. Particularly, if the data files are cached effectively and closer to users, it can significantly alleviate the overall network load during the high traffic DP. 
\end{itemize}

\subsection{Decoding Techniques} 
\label{decode}

\textcolor{black}{
Let us define STP as the average likelihood for the cache nodes, i.e., $C_1$, $C_2$, to successfully decode their intended files given by \cite{9475988},
\begin{equation}\label{stp}
\text{STP} = \frac{1}{2} \left(\Pr\left(C_1\text{ successful}\right) + \Pr\left(C_2\text{ successful}\right)\right).
\end{equation}
 We use four \textcolor{black}{different} decoding techniques in order to maximize the STP of the network by
optimizing the power split parameter, i.e., $\beta$ in \eqref{eqs}, and beamforming vectors, $\boldsymbol{v_{1}}, \boldsymbol{v_{2}}$ and $\boldsymbol{v_{1,2}}$ in \eqref{eq1}-\eqref{eq4}. The four techniques have their own levels of compromises between performance and complexity, suiting various use case requirements.} 

Here, two cache nodes are considered without the loss of generality, and thus, the results can be extended for multiple cache node cases.

\subsubsection{\textcolor{black}{Method 1. Joint Decoding with Successive Interference Cancellation (SIC)}}

Joint decoding with SIC allows efficient decoding of multiple simultaneous data streams in situations where cross-interference is present. Here, maximum ratio combining (MRC) and SIC are used during the DP to decode the cached sub-files received during the PP. The network STP is then given by \cite{9475988}, 

\begin{equation}
\text{STP} = \frac{1}{2}(\mu_1\gamma_1 + \mu_2\gamma_2).
\label{jointdec}
\end{equation}

Here, $\mu_1$ and $\mu_2$ are the probabilities of successfully decoding $B_1(t)$ and $A_2(t)$ respectively. Moreover, $\gamma_1$ denotes the probability of successfully decoding file $A$ after removing the interference $B_1(t)$ from the received signal. Similarly, $\gamma_2$ stands for the probability of successfully decoding $B(t)$ after removing the interference $A_2(t)$ using SIC.

\subsubsection{\textcolor{black}{Method 2. Joint Decoding without SIC}}

In this technique, the sub-files are decoded by assuming the interference as noise without using SIC. Thereby, the technique can be useful in cases where latency and complexity needs to be reduced. The STP is thereby expressed as \cite{9475988}

\begin{equation}
\text{STP} = \frac{1}{2}(\gamma_1 + \gamma_2),
\end{equation}

where $\gamma_1$ refers to the probability of decoding both $A_1(t)$ received during the PP and $A_2(t)$ received during the \textcolor{black}{high traffic} period. Similarly, $\gamma_2$ denotes the probability of decoding both $B_2(t)$ received during the \textcolor{black}{low traffic} period and $B_1(t)$ received during the high traffic period.

\subsubsection{Method 3. Separate Decoding with SIC}

In this technique, intended sub-files are decoded separately during the PP and DP. The network STP is given by \cite{9475988}

\begin{equation}
\text{STP} = \frac{1}{2}(\mu_1 \gamma_{11} \gamma_{12} + \mu_2 \gamma_{21} \gamma_{22}),
\label{seperatedec}
\end{equation}

where $\gamma_{11}$ denotes the probability of cache node 1 successfully decoding $A_1(T)$ during the PP while $\gamma_{22}$ denotes the probability of cache node 2 successfully decoding $B_2(t)$ during the PP. Moreover, $\gamma_{12}$ denotes the probability of cache node 1 successfully decoding sub-file $A_2(t)$ during the DP after decoding $B_1(t)$. In a similar context, $\gamma_{21}$ is the probability of cache node 2 decoding $B_1(t)$ after decoding $A_2(t)$. \textcolor{black}{Also,} $\mu_1$ and $\mu_2$ are the same as in \eqref{jointdec}.  In contrast to joint decoding, here, the data streams are decoded independently from each other.

\subsubsection{\textcolor{black}{Method 4.} Separate Decoding without SIC}

This technique decodes the sub-files seperately during the PP and DP without using the SIC. Thereby, the technique reduces the delay and complexity as the interference is considered as noise. 

Here, the network STP is given by \cite{9475988}

\begin{equation}
\text{STP}=\frac{1}{2}(\gamma_{11} \gamma_{12} + \gamma_{21} \gamma_{22}),
\end{equation}

where $\gamma_{11}$ and $\gamma_{22}$ are same as in \eqref{seperatedec}. Here, the probability of cache node 1 successfully decoding $A_2(t)$ by assuming $B_1(t)$ as noise is denoted by $\gamma_{12}$, while the probability of cache node 2 successfully decoding $B_1(t)$ by assuming $A_2(t)$ as noise is denoted by $\gamma_{21}$.

The optimization of the beams for networks having beamforming along
with CC scheme is done using the GA. The power split parameter $\beta$, and beamforming weights are optimized in order to maximize the STP, 
\textcolor{black}{
\begin{equation}
\max_{\boldsymbol{v_1}, \boldsymbol{v_2}, \boldsymbol{v_{1,2}}, \beta} \left(\mathrm{STP}\right) \quad s.t.
\end{equation}
\begin{align}
\beta \in {[0,1]}, \
|\boldsymbol{v_1}|^2 \leq 1,\
|\boldsymbol{v_2}|^2 \leq 1,\
|\boldsymbol{v_{1,2}}|^2 \leq 1.\
\end{align}}

 The system is power-limited by total transmit power $P$, and the optimization of power split parameter $\beta$ is carried out by a simple exhaustive search while GA is used to optimize the beams. Note that the beams of PP, and DP are optimized separately due to their non-simultaneous occurrence, which makes it unfeasible to optimize jointly.  

\begin{algorithm}
\label{alg}
\caption{Beamforming Optimization Algorithm}\label{alg:beamforming}
\begin{algorithmic}[1]
\STATE \textcolor{black}{Initialization: Randomly select $S$ set of beams from the pre-defined DFT-based codebook.}
\STATE Selection: For every precoding/beamforming matrix, calculate the objective metric, i.e., the minimum of the \textcolor{black}{signal-to-interference-plus-noise ratio (SINR)} observed by different cache nodes.
\STATE Select the best beamforming matrix called as the 'Queen' that yields the best result for the objective metric.
\STATE \textcolor{black}{Regeneration:} Save the Queen.
\textcolor{black}{
\STATE \textcolor{black}{Mutation:} Generate $J<S$ set of new beamforming matrices around the Queen, i.e., by choosing beams neighbouring the Queen. 
\STATE \textcolor{black}{Diversity:} The rest of the columns in the beamforming matrix are filled with $S-J-1$ set of new randomly picked beams from the pre-defined DFT-based codebook.
\STATE Proceed to Step 2 and continue the process for $N_{\text{it}}$ iterations pre-considered by the network designer.
\STATE Return the Queen as the final beamforming solution in the considered time slot.}
\end{algorithmic}
\end{algorithm}

Algorithm 1 gives the details of the GA-based beamforming optimization process. \textcolor{black}{Specifically}, the algorithm is based on the procedure detailed below. The algorithm is started by selecting $S$ possible random set of beams from the predefined Discrete Fourier Transform (DFT)-based codebook which is a collection of possible beamforming vectors. Here, the beamforming vector is characterized by its transmit power and phase shifts across the antennas facilitating to direct the signal towards specific direction. 
Then, in each iteration we evaluate the beamforming matrix with respect to the objective metric, i.e., minimum of SINR. The goal is to find the beamforming matrix \textcolor{black}{that yields the best result for the objective metric  observed by cache nodes.}
 Following the evaluation, the algorithm proceeds to identify the best performing beamforming matrix, the Queen, that results in the best result of the objective metric. 
Next, the algorithm generates $J<S$ new set of beamforming matrices by slightly modifying the Queen identified in the above step. In this step, minor modifications are made to the Queen, i.e., a local search is performed to explore the near-optimal solutions around the Queen in order to provide a balance between exploitation and exploration which eventually helps to achieve a global optimum. Then, the remaining columns in the beamforming matrix are filled with $S-J-1$ beamforming vectors randomly selected from the DFT codebook. This process avoids getting trapped in a local minimum prematurely as the search space remains diverse. 
This process will continue for $N_{\text{it}}$ iterations decided by the network designer. Finally, the Queen is returned as the optimum beamforming solution for the considered time slot. By optimizing the beamforming solution for each time slot independently, the algorithm serves the dynamics of the communication environment.

\section{Simulation Results And Discussion}
\label{simres}
In this section, we evaluate the proposed techniques with the performance metrics of our interest, i.e., STP and throughput. The simulation outcomes are produced for a network consisting of a single server, equipped with either multiple or a single transmit antenna, communicating to two cache nodes over Rayleigh fading channels. \textcolor{black}{The results are presented for different coded and uncoded caching schemes and different decoding methods}. The parameters adopted for the simulation configuration are listed in Table \ref{parameters}.

\begin{table}[h]
\centering
\caption{Simulation Parameters}
\begin{tabular}{|l|l|}
\hline
\textbf{Parameter} & \textbf{Value} \\
\hline
Number of transmit antennas at the server ($L$) & 32 \\
\hline
Number of cache nodes ($K$) & 2 \\
\hline
Number of channel realizations & 15000 \\
\hline
GA iterations & 150 \\
\hline
Predefined data rates & 2 npcu\\
\hline
\end{tabular}
\label{parameters}
\end{table}


The STP and throughput computations are performed for networks incorporating CC and uncoded caching schemes with the parameters listed above.

\begin{figure}
\centerline{\includegraphics[width=3.5in]{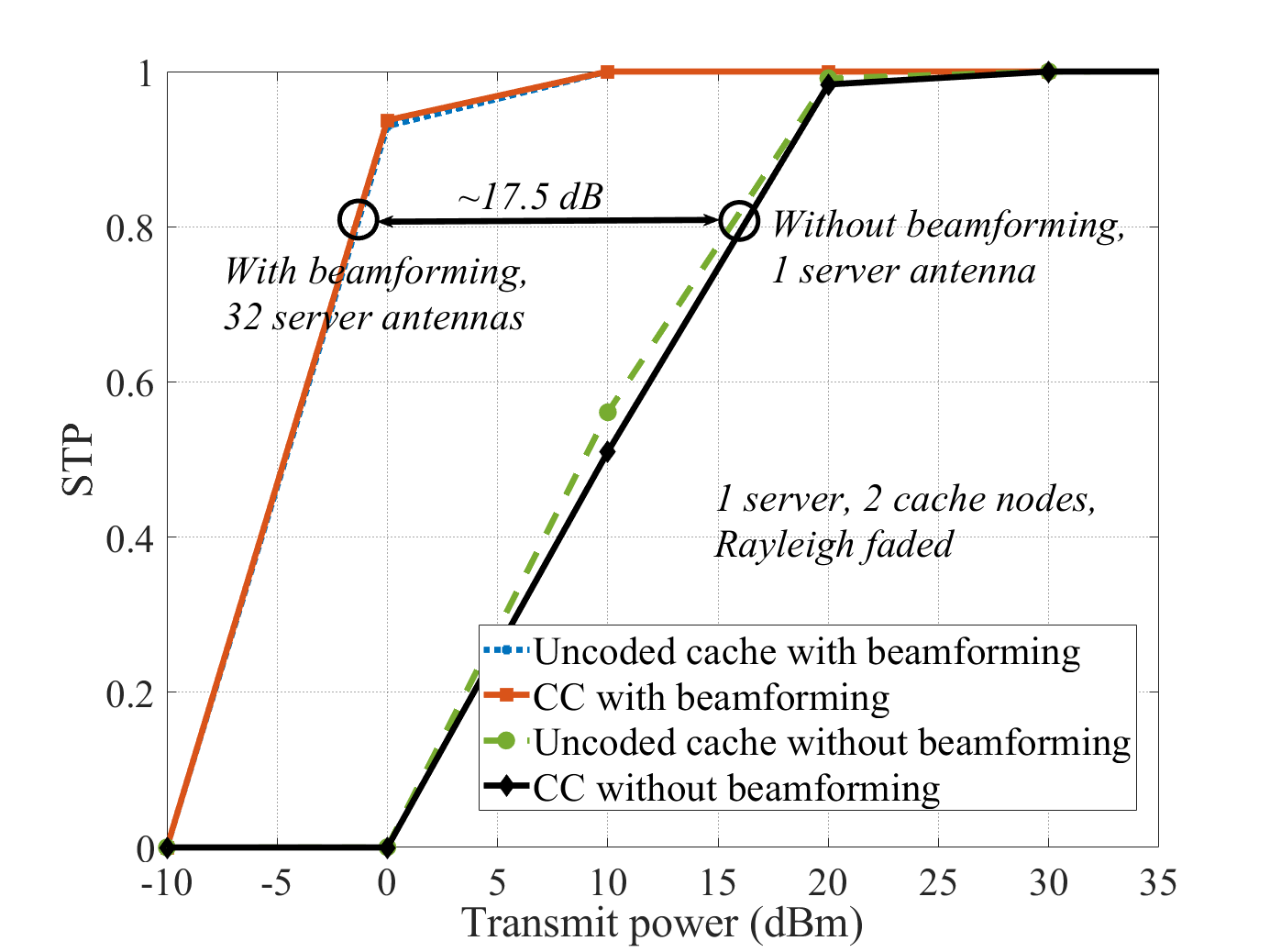}}
\caption{Network STP as a function of the transmit power in CC and uncoded caching schemes with and without beamforming with data rate of 2 nats-per-channel-use (npcu).}
\label{stp_beamform}
\end{figure}

In Fig. \ref{stp_beamform}, we compare the network STP as a function of transmit power for networks incorporating CC and uncoded caching. Here, for the CC scheme, we evaluate the performance of the joint decoding method without SIC in system with 2 single-antenna cache nodes, and 32 antennas at the server. \textcolor{black}{For cases without beamforming, the server consists of a single antenna.} As we see, both CC, and uncoded schemes with beamforming achieve significantly higher gains compared to a similar network without beamforming. Also, Fig. \ref{stp_beamform} shows that the network using an uncoded caching scheme performs marginally better than similar setups with coded caching with respect to STP. This is due to the fact that CC has to deal with varying inter cache node-server distance and interferences. However, the CC scheme overcomes delays in the network. Since the uncoded caching scheme signals are relatively more focused, they have better performance in terms of STP, but compromises the throughput.

\begin{figure}
\centerline{\includegraphics[width=3.5in]{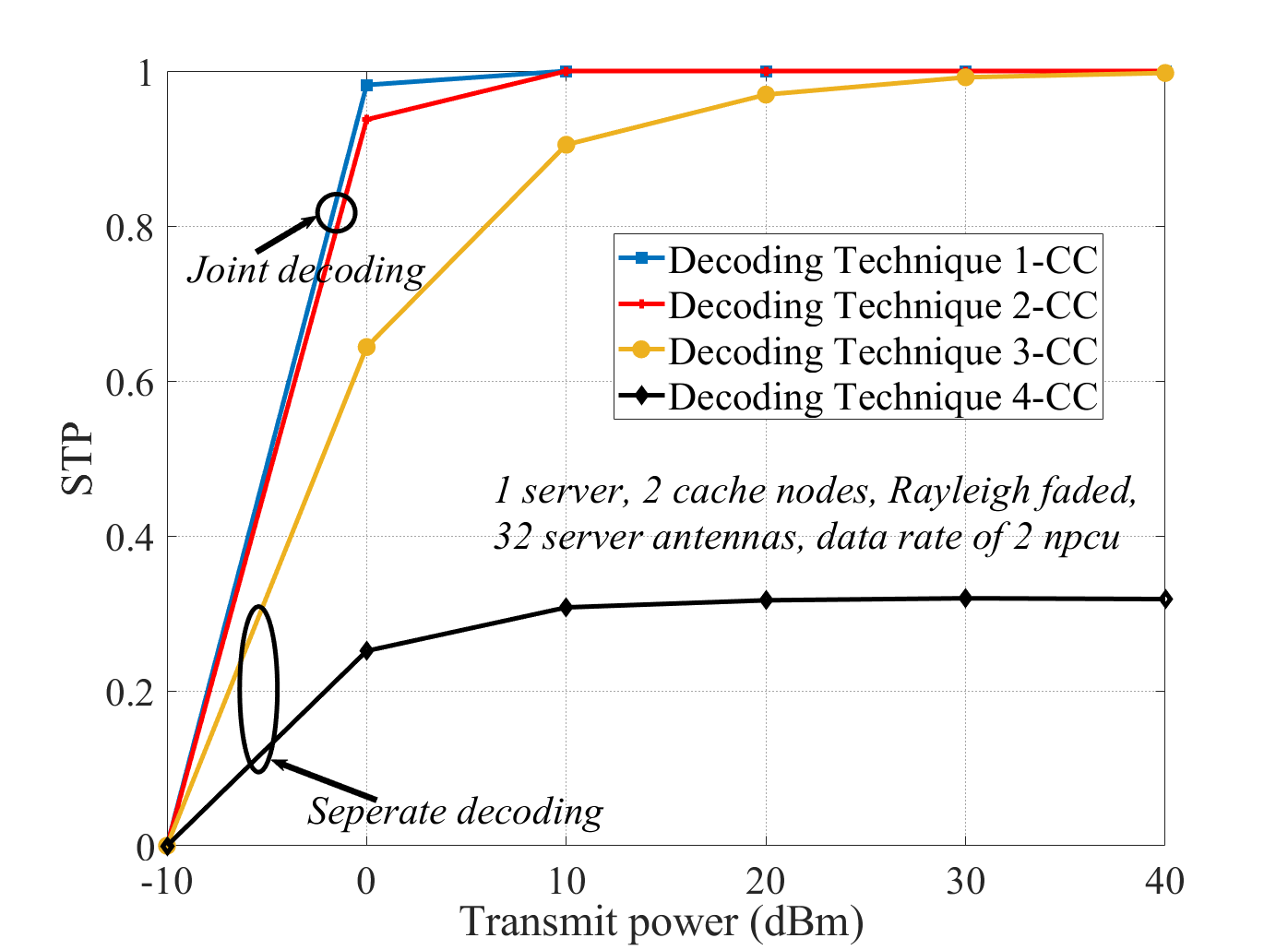}}
\caption{Network STP as a function of the the transmit power i with beamforming in a network with 2 cache nodes, 32 antennas at the server and data rate of 2 npcu.}
\label{stp_decoding}
\end{figure}

Figure \ref{stp_decoding} demonstrates the STP as a function of transmit power for a network with CC scheme in cases with the different decoding methods described in Section \ref{systemmodel}. Here, a setup with single server with 32 antennas, 2 \textcolor{black}{single-antenna} cache nodes and a data rate of 2 npcu is considered. As seen in Fig. \ref{stp_decoding}, the performance of decoding methods 1 and 2, i.e., joint decoding with SIC and joint decoding without SIC, respectively, are closely aligned. However, here, joint decoding with SIC performs slightly better than the latter. As observed, there is a notable difference in performance between the methods, separate decoding with SIC and separate decoding without SIC. A similar performance can be observed in the scenarios without beamforming. This difference suggests that the joint decoding of sub-files significantly enhances the network STP, in comparison to individual decoding. However, it should also be mentioned that the performance increase in joint decoding methods also raises the decoding complexity. In contrast, decoding Methods 3 and 4, i.e., seperate decoding with SIC and seperate decoding without SIC methods present a drawback. In these methods, cache nodes may decode sub-files during the low traffic period that are not subsequently demanded by users during the high traffic period. Nevertheless, when utilized in combination with appropriate beamforming and joint decoding, the impact of SIC-based interference cancellation on network STP is minimal. This accounts for the negligible difference in the STP between joint decoding with and without SIC. When sub-files are decoded separately, the significance of interference cancellation is increased, resulting in a notable improvement of STP for decoding method 3 as compared to method 4.

\begin{figure}
\centerline{\includegraphics[width=3.5in]{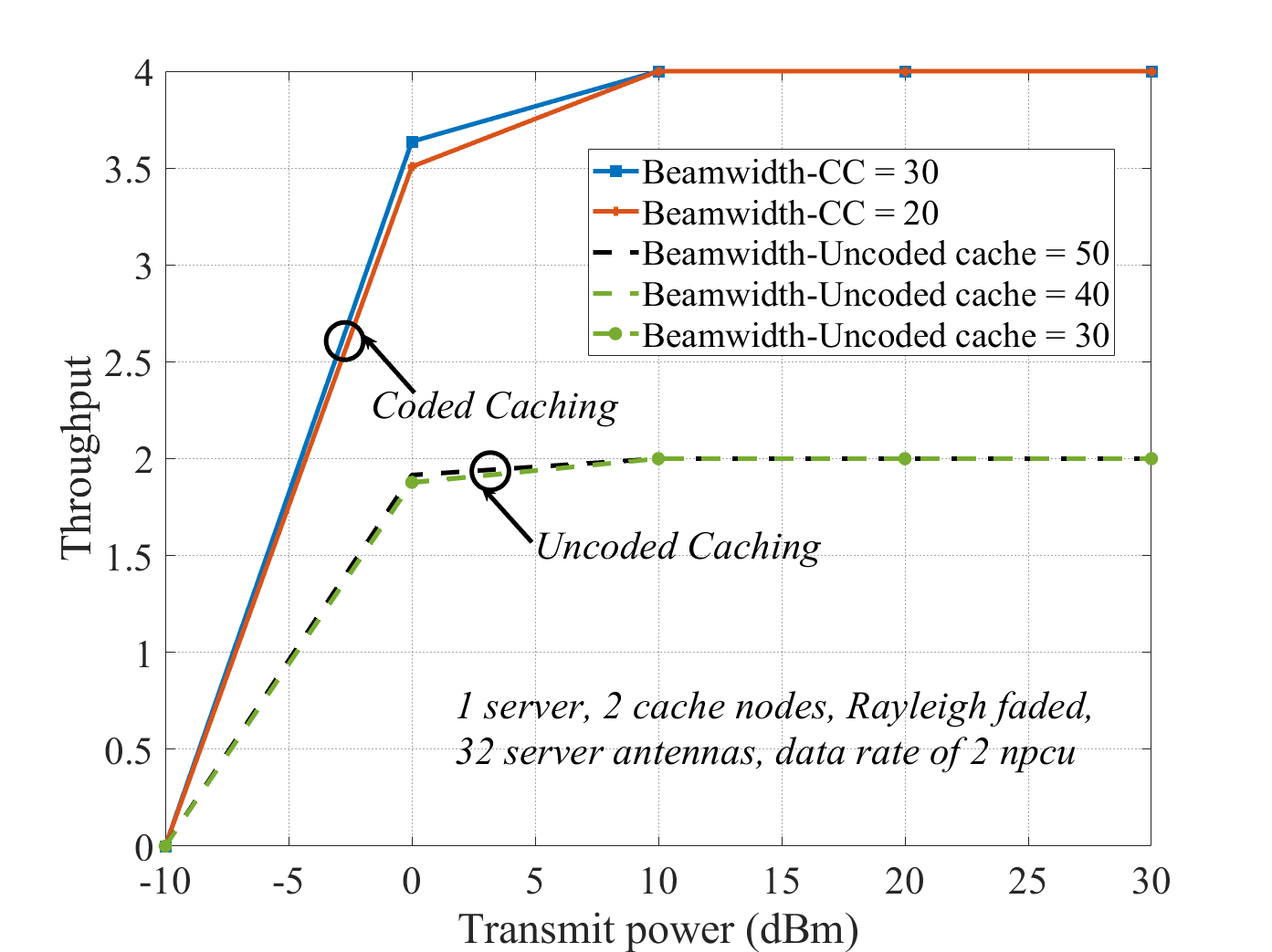}}
\caption{Network throughput as a function of the transmit power in CC and un-coded caching
schemes with beamforming with different beamwidths for a network with 2 \textcolor{black}{single-antenna} cache
nodes, 32 antennas at the server and data rate of
2 npcu.}
\label{beamwidth}
\end{figure}

Figure \ref{beamwidth} compares the network throughput considering both CC and uncoded caching schemes together with beamforming with varying quantities of DFT-based beams. \textcolor{black}{Here, for the CC scheme, the throughput is given by 
\begin{align}
\eta=D_1 \Pr\left(C_1\text{ successful}\right)
+D_2 \Pr\left(C_2\text{ successful}\right),
\end{align}
where $D_i,i = 1,2$, denote the date rate at $i^{\text{th}}$ cache node, while $\Pr\left(C_1\text{ successful}\right)$
is the probability of successful decoding in cache node 1. Similarly, $\Pr\left(C_2\text{ successful}\right)$ represents the probability of successful decoding in cache node 2. In this scheme, we use the decoding techniques described in Section \ref{decode}}.

\textcolor{black}{
Moreover, for the uncoded caching scheme the throughput can be expressed as 
\begin{align} \label{deceq}
\eta=\frac{1}{2}(D_1 \Pr\left(C_1'\text{ successful}\right)
+D_2 \Pr\left(C_2'\text{ successful}\right)).
\end{align}
Here, $\Pr\left(C_1'\text{ successful}\right)$ and $\Pr\left(C_2'\text{ successful}\right))$ denote the probabilities of successful decoding in cache node 1 and cache node 2 respectively. The signal transmission happens in two time slots during the high traffic period unlike the CC scheme and thus, the throughput gets divided by a factor of two as denoted in \eqref{deceq}}.


We use 2 cache nodes, a single server with 32 antennas and a data rate of 2 npcu. A considerable throughput increase is observed in the network deploying CC with beamforming compared to the network implementing uncoded caching with beamforming. In particular, despite the slight reduction of STP associated with the CC scheme compared to the uncoded caching scheme, CC significantly mitigates backhaul traffic during high traffic periods.

\begin{figure}
\centerline{\includegraphics[width=3.5in]{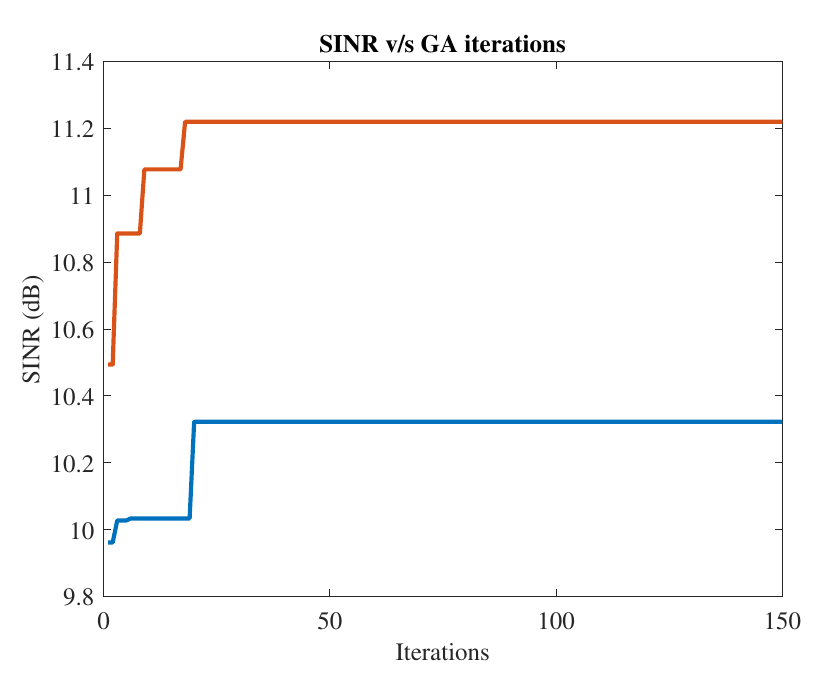}}
\caption{Minimum SINR of the cache nodes as a function of the GA iterations for a network with 2 cache nodes, 32 antennas at the server, $P = 60$ dBm and 150 GA iterations.}
\label{gatiteration}
\end{figure}

Figure \ref{gatiteration} provides a series of examples illustrating the convergence behavior of the GA in Algorithm 1. The results represent the minimum SINR registered by both cache nodes, as a function of number of interations. This setup considers a configuration of two cache nodes, single server equipped with 32 antennas, a transmit power of $P$ = 60 dBm, and a total of 150 GA iterations. As we see, the GA demonstrates rapid convergence after a relatively limited number of iterations. However, there are instances where the GA becomes transiently trapped in a local minima. However, due to the implementation of step 5 in Algorithm 1, the GA is equipped with the ability to free itself from the local minima, ultimately converging towards the global optimal solution.

\section{Conclusion}
\textcolor{black}{
We studied the influence of adaptive transmission, variety of decoding methods, and distinct caching schemes on networks employing beamforming together with the coded caching scheme. The findings indicate that by using caching and beamforming together, backhaul traffic can be minimized and network STP can be maximized. Consequently, interference and delays within the network are diminished, leading to enhanced gains and improved network throughput. In addition, we developed an effective GA-based method for optimizing beamforming in CC networks, demonstrating promising outcomes in reducing backhaul traffic.}

\section*{Acknowledgment}

This work was supported in part by the Gigahertz-ChaseOn Bridge Center at Chalmers in a project financed by Chalmers, Ericsson, and Qamcom.


\bibliographystyle{IEEEtran}
\bibliography{bibliography}

\end{document}